\documentclass[%
 aip,
 amsmath,amssymb,
 reprint,%
]{revtex4-1}

\usepackage{graphicx}
\usepackage{dcolumn}
\usepackage{bm}
\usepackage{caption}
\usepackage{subcaption}
\usepackage{xcolor}

\usepackage[utf8]{inputenc}
\usepackage[T1]{fontenc}
\usepackage{mathptmx}
\usepackage{etoolbox}

\newcommand{\iotabar}{\iota\!\!\!\text{-}}
\makeatletter
\def\@email#1#2{%
 \endgroup
 \patchcmd{\titleblock@produce}
  {\frontmatter@RRAPformat}
  {\frontmatter@RRAPformat{\produce@RRAP{*#1\href{mailto:#2}{#2}}}\frontmatter@RRAPformat}
  {}{}
}%
\makeatother
\begin{document}

\preprint{AIP/123-QED}

\title[]{Stellarator optimization for nested magnetic surfaces at finite $\beta$ and toroidal current}
\author{A. Baillod}
 \email{antoine.baillod@epfl.ch}
\author{J. Loizu}
\author{J. P. Graves}%
\affiliation{ 
\'Ecole Polytechnique F\'ed\'erale de Lausanne, Swiss Plasma Center, CH-1015 Lausanne, Switzerland
}%

\author{M. Landreman}
\affiliation{
Institute for Research in Electronics and Applied Physics, University of Maryland, College Park MD 20742, USA
}%

\date{\today}

\begin{abstract}
Good magnetic surfaces, as opposed to magnetic islands and chaotic field lines, are generally desirable for stellarators. In previous work, M. Landreman \textit{et al.} [Phys. of Plasmas \textbf{28}, 092505 (2021)] showed that equilibria computed by the Stepped-Pressure Equilibrium Code (SPEC) [S. P. Hudson \textit{et al.}, Phys. Plasmas \textbf{19}, 112502 (2012)] could be optimized for good magnetic surfaces in vacuum. In this paper, we build upon their work to show the first finite-$\beta$, fixed- and free-boundary optimization of SPEC equilibria for good magnetic surfaces. The objective function is constructed with the Greene's residue of selected rational surfaces and the optimization is driven by the SIMSOPT framework [M. Landreman \textit{et al.}, J. Open Source Software \textbf{6}, 3525 (2021)]. We show that the size of magnetic islands and the consequent regions occupied by chaotic field lines can be minimized in a classical stellarator geometry by optimizing either the injected toroidal current profile, the shape of a perfectly conducting wall surrounding the plasma (fixed-boundary case), or the coils (free-boundary case), in a reasonable amount of computational time. This work shows that SPEC can be used as an equilibrium code both in a two-step or single-step stellarator optimization loop.
\end{abstract}

\maketitle

\section{\label{sec:intro}Introduction}

In toroidal geometries, three dimensional (3D) magnetohydrodynamic (MHD) equilibria are, in general, a mix of nested magnetic surfaces, magnetic islands and magnetic field line chaos \cite{Helander2014a, Hanson1984EliminationStellarators, Cary1986}.
In the plasma core, the latter two topologies are usually detrimental to confinement, \textit{i.e.} the radial transport of particles and energy is generally greater than in regions of nested magnetic surfaces.
In addition to other desirable properties, a common target of stellarator optimization is to increase the volume occupied by magnetic surfaces\cite{Hudson2002}.


The equilibrium approach for optimizing the volume occupied by nested flux surfaces requires three tools; (i) a fast 3D equilibrium code that does not assume nested flux surfaces, (ii) a numerical diagnostic that provides a measure of integrability of the magnetic field \cite{Meiss1992SymplecticTransport,MacKay1985ConversePractice,Greene1978,Loizu2017}, and (iii) an optimization algorithm.
Two fast 3D equilibrium codes are the Variational Moments Equilibrium Code\citep{Hirshman1983Steepest-descentEquilibria,Hirshman1986} (VMEC) and the Stepped Pressure Equilibrium Code\citep{Hudson2012b} (SPEC). VMEC is based however on the assumption of the existence of nested flux surfaces everywhere and is thus, when used as a stand-alone code, not suitable for the optimization of magnetic islands and chaotic field lines. On the other hand, SPEC does not assume the existence of nested flux surfaces everywhere in the plasma. 
In addition, SPEC has been recently extended to allow free-boundary calculations \citep{Hudson2020c}, and to allow the prescription of net toroidal current profiles \citep{Baillod2021}. Numerical work has also improved its robustness and speed \citep{Qu2020}.


Most physics codes developed by the stellarator community are based on the assumption of nested flux surfaces and thus require a VMEC equilibrium as input. In a recently published paper, \citet{Landreman2021a} showed that by optimizing the plasma boundary, SPEC can be used in combination with VMEC to obtain self-consistent vacuum configurations where both codes are in agreement, ensuring good magnetic surfaces in the region of interest. This allows then to trust in any auxiliary codes that assume nested flux surfaces in this configuration, and to safely optimize for other metrics. 

In this paper, we extend the work by \citet{Landreman2021a} by showing that finite-$\beta$ SPEC equilibria with non-zero net toroidal currents can also be optimized to reduce the volume occupied by magnetic islands and field line chaos in a reasonable amount of time. Additionally, we explore the use of parameter spaces other than the plasma boundary that could be of interest. Indeed, we leverage new capabilities of SPEC to show that the volume of magnetic surfaces in a stellarator can be maximized by optimizing the injected toroidal current profile, or the coil configuration --- two experimentally relevant "knobs" \cite{Geiger2015}. For the optimization, we follow \citet{Landreman2021a} and use the SIMSOPT framework \cite{Landreman2021SIMSOPT:Optimization}, which in particular can construct an objective function based on Greene's residues \cite{Greene1978} of some selected rational surfaces.

\section{\label{sec:method}Method}

Since both fixed- and free-boundary equilibria optimization are considered in this paper, we first describe how these are computed by SPEC. 

SPEC computes equilibria where the pressure profile is stepped, \textit{i.e.} a finite number of nested surfaces support a pressure discontinuity. Between these interfaces, there is a finite volume where the pressure is constant and the force-free magnetic field is described by a Taylor state \citep{Taylor1974,Taylor1986}. These interfaces thus describe a set of $N_{vol}$ nested annular regions, called \emph{SPEC volumes} or simply \emph{volumes} in what follows. Details about SPEC can be found in \citet{Hudson2012b,Hudson2020c} and references therein. 

Using the standard cylindrical coordinate system $(R,\phi,Z)$, a plasma boundary surface $\Gamma_{PB}$ is parametrized by $R(\theta,\phi),\ Z(\theta,\phi)$, where $\theta$ is a poloidal angle. A fixed-boundary SPEC equilibrium is then determined by $\Gamma_{PB}$, the number of volumes $N_{vol}$ and the toroidal flux they enclose $\{\psi_{l}\}_{l=\{1,\ldots,N_{vol}\}}$, the pressure in each volume $\{p_l\}_{l=\{1,\ldots,N_{vol}\}}$, the net toroidal  current flowing at the volumes' interfaces $\{I^s_{l,\phi}\}_{l=\{1,\ldots,N_{vol}\}}$ and the net toroidal current flowing in each volume $\{I^v_{l,\phi}\}_{l=\{1,\ldots,N_{vol}\}}$.
Interface currents represent all equilibrium pressure-driven currents, such as diamagnetic, Pfirsch-Schl\"uter, and bootstrap currents, as well as shielding currents arising when an ideal interface is positioned on a resonance \citep{Loizu2015}, while volume currents represent externally driven currents such as Electron Cyclotron Current Drive (ECCD), Neutral Beam Current Drive (NBCD) or Ohmic current \citep{Baillod2021}. Note that SPEC can be run with different inputs --- this will  however not be covered in the present paper.

A free-boundary equilibrium is determined by a computational boundary surface $\Gamma_{CB}$ surrounding the plasma which is enclosed by the coils, and the Fourier harmonics $V_{mn}$ of the component of the vacuum magnetic field normal to $\Gamma_{CB}$, \textit{i.e.}
    
\begin{equation}
    \mathbf{B}^c\cdot\mathbf{n} = \sum_{n=-N_{tor}}^{N_{tor}}\sum_{m=0}^{M_{pol}} V_{mn} \sin(m\theta-nN_{fp}\phi),
\end{equation}
with $\mathbf{n}=\mathbf{e}_\theta\times \mathbf{e}_\phi$ a vector normal to $\Gamma_{CB}$, $\mathbf{B}^c$ the magnetic field produced by the coils, $M_{pol}$ and $N_{tor}$ are respectively the number of poloidal and toroidal Fourier modes used in the calculation, $N_{fp}$ is the number of field periods, and stellarator symmetry has been assumed. The Fourier harmonics $V_{mn}$ can be obtained from the coil geometry and current; changing these harmonics is equivalent to changing the coil system. In addition, the calculation of a free-boundary equilibrium requires specifying the total toroidal current flowing in the plasma, the total current flowing in the coils, and as in fixed-boundary the profiles $\{\psi_{l}\}_{l=\{1,\ldots,N_{vol}\}}$,  $\{p_l\}_{l=\{1,\ldots,N_{vol}\}}$,  $\{I^s_{l,\phi}\}_{l=\{1,\ldots,N_{vol}\}}$ and $\{I^v_{l,\phi}\}_{l=\{1,\ldots,N_{vol}\}}$.

We now consider the optimization of a finite-$\beta$, finite current, classical stellarator equilibrium constructed with SPEC, which presents regions of magnetic islands and magnetic field line chaos. A classical stellarator geometry was chosen for simplicity, as few Fourier modes are required to described the equilibrium. An experimental instance of a classical stellarator was W7-A \cite{Grieger1985}. However, the optimization procedure presented here does not depend on the specific choice of geometry. 

We construct a free-boundary equilibrium which will be the initial state for the optimizations presented in this paper. We choose the computational boundary $\Gamma_{CB}$ to be a rotating ellipse, 

\begin{align}
    R^{CB}(\theta,\phi) &= R^{CB}_{00} + R^{CB}_{10}\cos\theta + R^{CB}_{11}\sin(\theta-N_{fp}\phi), \label{eq.rotellipse1}\\
    Z^{CB}(\theta,\phi) &= Z^{CB}_{10}\sin\theta + Z^{CB}_{11}\sin(\theta-N_{fp}\phi),\label{eq.rotellipse2}
\end{align}
where $R^{CB}_{00}=10$m, $R^{CB}_{10}=-Z_{10}=1$m, and $R^{CB}_{11}=Z_{11}=0.25$m. We impose a pressure profile linear in toroidal flux, that we approximate with seven steps, namely $N_{vol}=8$, with the last volume being a vacuum. We assume no externally induced currents by setting the total toroidal current flowing in each volume to zero, $\{I^v_{\phi,l}\}=0$, and we assume that the plasma generates a bootstrap-like toroidal current proportional to the pressure jump at the interface  $I^s_{\phi,l}\propto (p_{l+1}-p_l)$ (see Figure \ref{fig:Iprofile}). Finally, we suppose the existence of coils such that $\mathbf{B}\cdot\mathbf{n}=0$ on $\Gamma_{CB}$ at a plasma average beta, $\langle\beta\rangle$, of $1.5\%$, where $\mathbf{B}=\mathbf{B}^c+\mathbf{B}^p$ and $\mathbf{B}^p$ is the magnetic field produced by the plasma currents. Note that the condition $\mathbf{B}\cdot\mathbf{n}=0$ is only valid for the specified $\Gamma_{CB}$, $\langle\beta\rangle$ and profiles. We will refer to this initial equilibrium as the \emph{free-boundary equilibrium}; its associated un-optimized magnetic topology is shown via its Poincare section and rotational transform, plotted on the top left panel of Figure \ref{fig:poincare_plots} and on Figure \ref{fig:iota_profile} respectively. The discontinuities observed in the rotational transform profile are due to SPEC stepped-pressure equilibrium model --- since the magnetic field is generally discontinuous across the interfaces, and so therefore is the rotational transform.

The exact same equilibrium can be generated with fixed-boundary conditions if we assume that a perfectly conducting wall $\Gamma_w$, parametrized by $R^w(\theta,\phi)$ and $Z^w(\theta,\phi)$, is positioned at the computational boundary given by Eqs.(\ref{eq.rotellipse1})-(\ref{eq.rotellipse2}). The difference is, however, that for any value of $\langle\beta\rangle$ and choice of  profiles $\{I^v_{\phi,l}\}$, $\{I^s_{\phi,l}\}$, the condition $\mathbf{B}\cdot\mathbf{n}=0$ would remain valid. We will refer to this equilibrium as the \emph{perfectly conducting wall equilibrium}. Note that the free-boundary equilibrium has effectively no conducting wall (no wall limit).

We now consider different degrees of freedom depending on the type of initial equilibrium. 
For the free-boundary equilibrium, the parameter space is a selected set of the Fourier harmonics $V_{mn}$, which emulate an optimization of the coil geometry and current, as would be done in a "single-step stellarator optimization" \cite{Hudson2002}.
For the perfectly conducting wall equilibrium, we consider two different parameter spaces. The first is the current flowing in the plasma volumes, $\{I^v_{l,\phi}\}_{l=\{1,\ldots,N_{vol}\}}$, which is the externally induced net toroidal plasma current. This parameter space is relevant for example for scenario design, where one could desire to heal magnetic islands and chaos for a given plasma geometry and coil system.  
The second parameter space is the geometry of the wall expressed as Fourier series,
\begin{align}
    R^w(\theta,\phi) &= \sum_{m=0}^{M_{pol}}\sum_{n=-N_{tor}}^{N_{tor}}R^w_{mn} \cos(m\theta-n N_{fp}\phi)\\
    Z^w(\theta,\phi) &= \sum_{m=0}^{M_{pol}}\sum_{n=-N_{tor}}^{N_{tor}}Z^w_{mn} \sin(m\theta-n N_{fp}\phi).
\end{align}
The degrees of freedom are then a selected set of Fourier harmonics $\{R^w_{mn},Z^w_{mn}\}$.

The objective functions for each optimization are based on Greene's residues \citep{Greene1978}. The Greene's residue $R$ is a quantity that can be computed for any periodic orbit, with $R=0$ when the island width is zero, $R< 1$ for an O-point, and $R>1$ for an X-point. The objective function is
\begin{equation}
    f_G = \sum_i R_{i}^2, \label{eq:objective_function_1}
\end{equation}
where $R_{i}$ is the residue for a field line on the  $i^{\textit{th}}$ targeted island chain. The downside of this approach is that each resonant field line has to be selected by hand before starting the optimization. If a new resonance becomes excited during the optimization, it will not be included in the objective function. From the initial equilibrium, Fig.\ref{fig:poincare_plots} (top left), we can identify a set of resonant rational surfaces and their associated residue, listed in Table \ref{tab:residues}. We use the SIMSOPT framework to drive the optimization, which is based on the default \emph{scipy.optimize} \citep{Virtanen2020} python algorithm for non-linear least squares optimization.

\begin{table}
\centering
\begin{tabular}{c c c c c c }
    Index & Volume $l$ & toroidal mode $n$ & poloidal mode $m$  & $\iotabar$    \\
     \hline
    1 & 8 & 5 & 9 & $0.55$ \\
    2 & 8 & 5 & 8 & $0.63$ \\
    3 & 8 & 5 & 7 & $0.71$ \\
    4 & 5 & 5 & 6 & $0.83$ \\
    5 & 5 & 5 & 5 & $1.00$ \\
    6 & 4 & 5 & 5 & $1.00$ \\
    7 & 3 & 5 & 6 & $0.83$ \\
    8 & 3 & 5 & 5 & $1.00$ \\
    9 & 3 & 5 & 4 & $1.25$ \\
\end{tabular}
\caption{Identified resonant surfaces and their rotational transform $\iotabar$ from the initial equilibrium.}
\label{tab:residues}
\end{table}

\section{Results}

\subsection{Free-boundary optimization}

We start by optimizing the free-boundary equilibrium. Residues 1-3 of Table \ref{tab:residues} are used to build the objective function according to Eq.(\ref{eq:objective_function_1}). These residues correspond to the islands and subsequent chaos present in the vacuum region, right outside the plasma edge. The parameter space is a selected set of Fourier modes $\{V_{mn}\}$,  $(m,n)=\{(6,n)\}_{n=\{-3,\ldots,3\}}$. More residues could be targeted if more $\{V_{mn}\}$ modes were used as degrees of freedom. This is however computationally expensive and was not done for this proof-of-principle calculation.

\begin{figure*}
\centering
\hfill
\begin{subfigure}[c]{0.45\textwidth}
    \centering
    \includegraphics[width=0.95\textwidth]{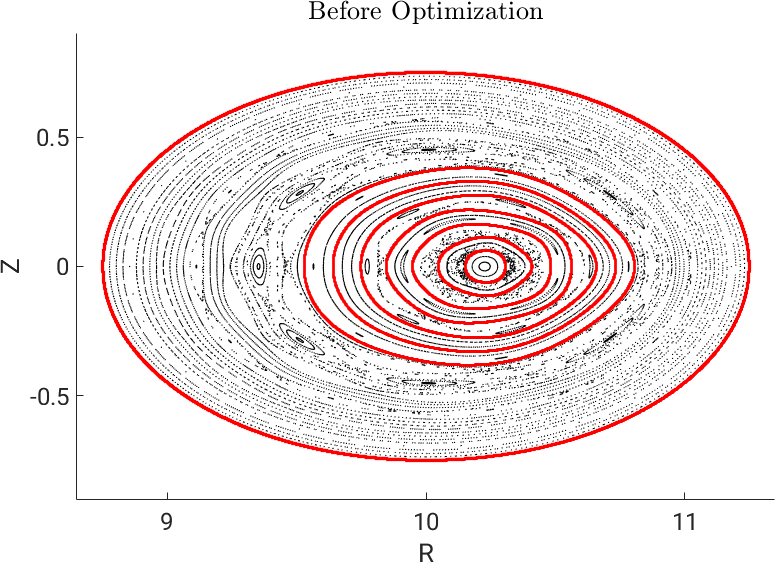}   
\end{subfigure}
\hfill
\begin{subfigure}[c]{0.45\textwidth}
    \centering
    \includegraphics[width=0.95\textwidth]{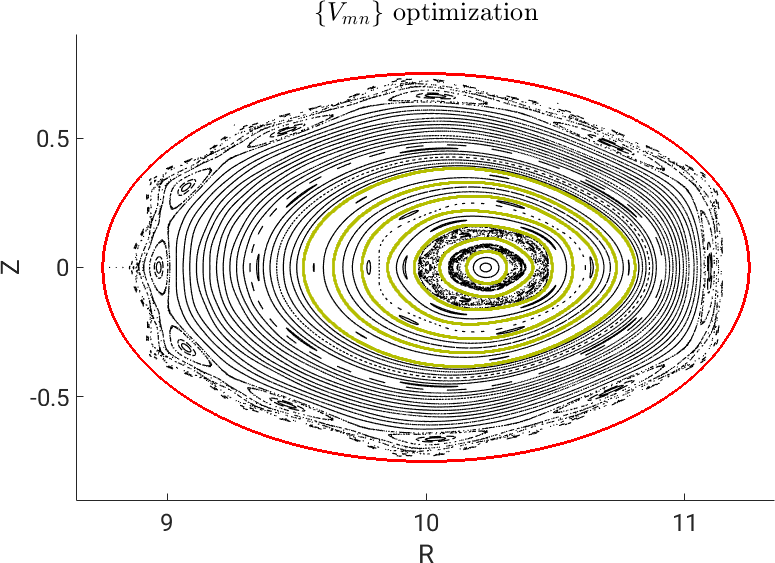}
\end{subfigure}
\hfill \\
\centering
\hfill
\begin{subfigure}[c]{0.45\textwidth}
    \centering
    \includegraphics[width=0.95\textwidth]{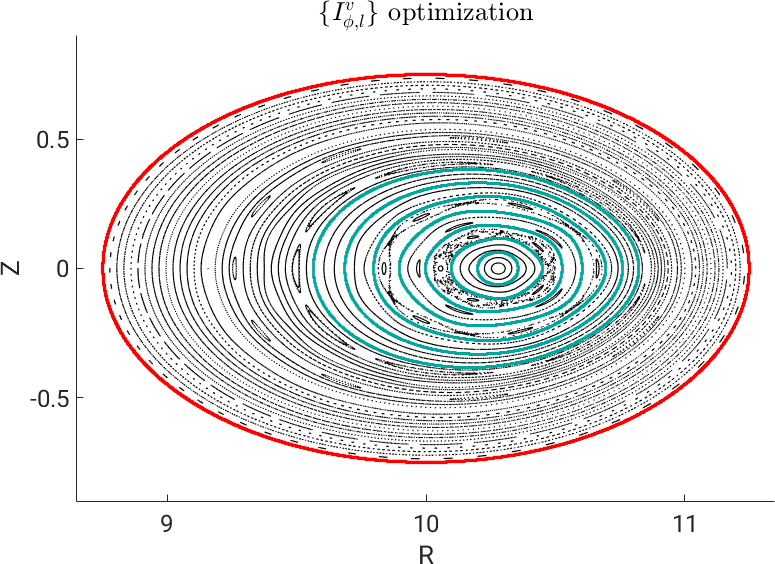}
\end{subfigure}
\hfill
\begin{subfigure}[c]{0.45\textwidth}
    \centering
    \includegraphics[width=0.95\textwidth]{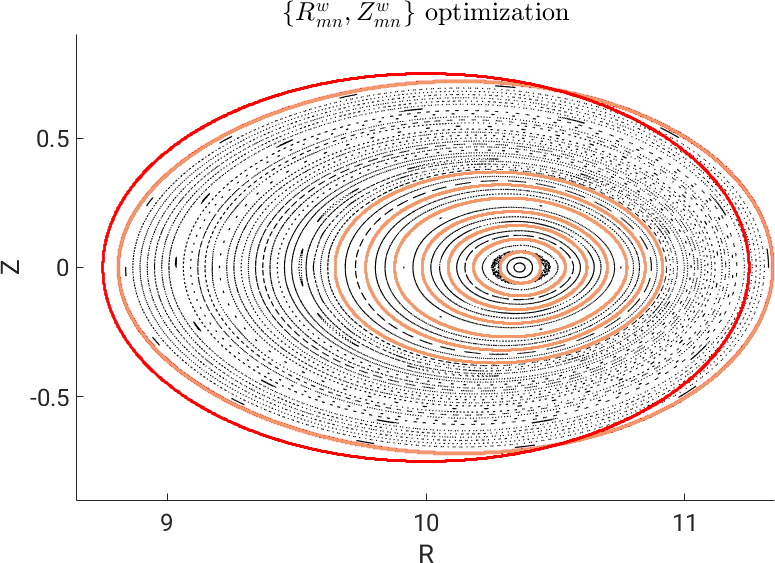}
\end{subfigure}
\hfill
\caption{Poincare section of the initial and optimized equilibria. Top left: Initial equilibrium (free-boundary and perfectly conducting wall cases). Red magnetic surfaces are SPEC's interfaces which support the pressure jumps. Outermost red interface is $\Gamma_{CB}$. Top right: free-boundary equilibrium, optimization of $\{V_{mn}\}$. In red: $\Gamma_{CB}$ is the same as the initial $\Gamma_{CB}$. In yellow:  inner interfaces of the optimized equilibrium. Bottom left: optimization of $\{I^v_{\phi,l}\}_{l=\{1,\ldots,7\}}$. In red: $\Gamma_w$ is the same as the initial equilibrium. In blue: inner interfaces of the optimized equilibrium. Bottom right: optimization of $\Gamma_w$. In orange, optimized $\Gamma_w$ and inner interfaces, in red, initial $\Gamma_w$. The bottom plots are fixed-boundary equilibria.   }
\label{fig:poincare_plots}
\end{figure*}

The Poincare section of the optimized equilibrium is plotted on the top right of Figure \ref{fig:poincare_plots}. As expected, $\Gamma_{CB}$ is no longer a flux surface since $\mathbf{B}^c$ has been changed relative to the un-optimized equilibrium shown on the top left panel of Figure \ref{fig:poincare_plots}. The difference between the magnetic field generated by the coils pre- and post-optimization is of the order of 1\% of the total magnetic field, \textit{i.e.} $\delta B/B \sim 1\%$. We observe that the targeted island chains have disappeared. However, new resonances appeared close to the computational boundary $\Gamma_{CB}$, in particular one with mode number $(m,n)=(10,5)$. The residues associated with these resonances are not included in the objective function, which explains why the optimizer converged to this state. Nevertheless, this optimization demonstrates that the parameter space $\{V_{mn}\}$ is suitable for stellarator optimization.

The rotational transform profile of the optimized equilibrium is plotted on Figure \ref{fig:iota_profile}. We observe that the rotational transform profile after optimization is of the same order as before the optimization --- it still crosses the same rationals, but these rationals do not resonate anymore!

\subsection{Perfectly conducting wall optimization}

 We now look at two different optimizations of the perfectly conducting wall equilibrium. In the first one, we only target the residues in the vacuum region, \textit{i.e.} residues 1-3 of Table \ref{tab:residues}, and the parameter space is the profile of current flowing in the plasma volumes $\{I^v_{\phi,l}\}_{l=\{1,\ldots,7\}}$. If more residues were included in the objective function, the optimizer could not lower the objective function sufficiently to observe an effect on the island width. This might be because not enough degrees of freedom were provided. Increasing the number of volumes in SPEC is however not a solution, since it adds additional topological constraints on the magnetic field and some island chains might remain undetected. Note that this optimization could also be achieved in free-boundary, but fixed-boundary calculations were considered here for simplicity. 
 
 In the second optimization, all residues listed in Table \ref{tab:residues} were included in the objective function and the geometry of the perfectly conducting wall $\Gamma_w$ was optimized. The selected degrees of freedom are the modes $R^w_{mn}$ and $Z^w_{mn}$ with $(m,n)=(1,1),\ (1,2),\ (2,1)$. 

Figure \ref{fig:poincare_plots} shows the result of the optimization of $\{I^v_{\phi,l}\}_{l=\{1,\ldots,7\}}$ (bottom left) and the optimization of $\Gamma_w$ (bottom right). Comparing both Poincare plots with the initial equilibrium (top left), we observe that the targeted residues have indeed been minimized - the islands are now much smaller, or even disappeared in some cases.

\begin{figure*}
    \centering
    \hfill
    \begin{subfigure}[c]{0.45\textwidth}
    \includegraphics[width=.95\textwidth]{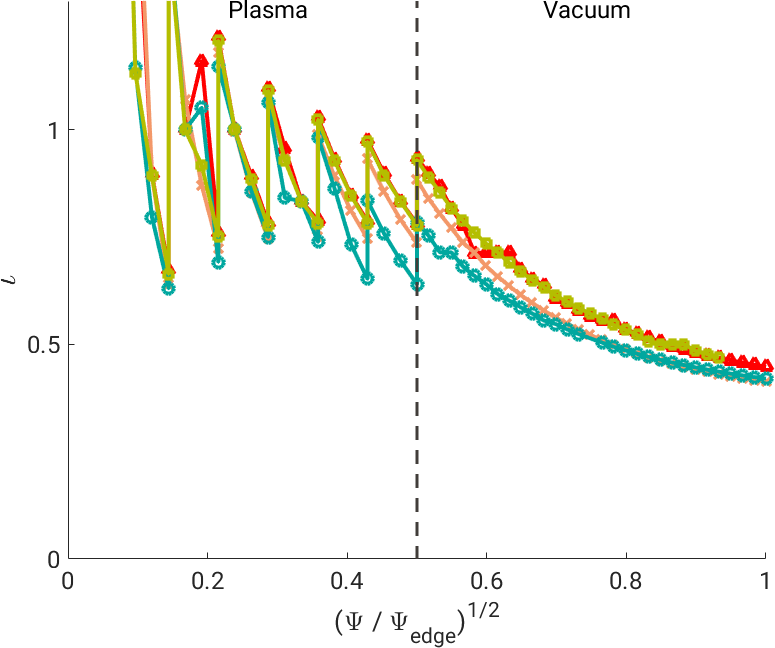}
    \end{subfigure}
    \hfill
    \begin{subfigure}[c]{0.45\textwidth}
    \includegraphics[width=.95\textwidth]{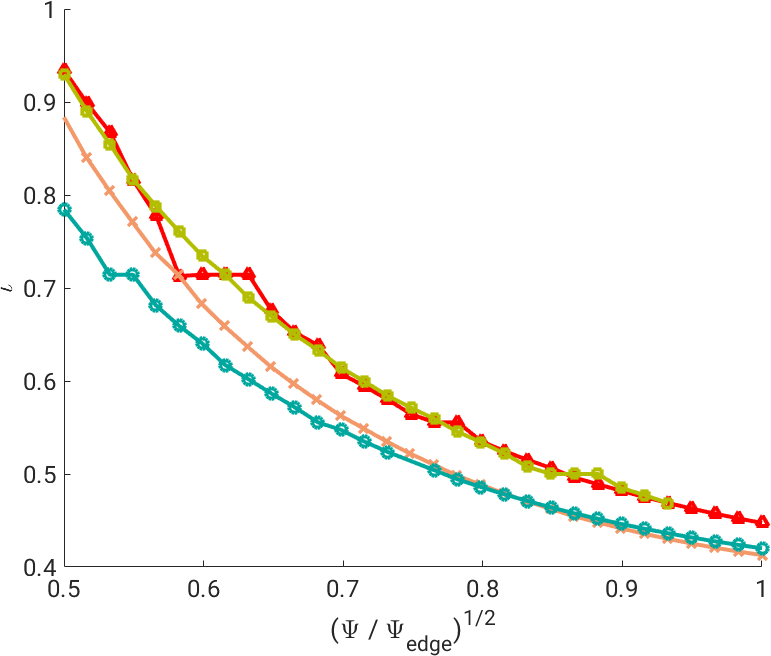}
    \end{subfigure}
    \hfill
    \caption{Rotational transform as a function of the normalized square-root of the toroidal flux $\Psi$, where $\Psi_{edge}$ is the toroidal flux enclosed by $\Gamma_{w}$. Left: Full profile, right: zoom on vacuum region. Red triangles: initial equilibrium and profiles. Orange crosses: optimization of $\Gamma_w$. Blue circles: Optimization of $\{I^v_{\phi,l}\}_{l=\{1,\ldots,7\}}$. Green squares: optimization of $\{V_{mn}\}$. Dashed, black line: position of the plasma-vacuum interface. }
    \label{fig:iota_profile}
\end{figure*}

\begin{figure}
    \centering
    \includegraphics[width=.45\textwidth]{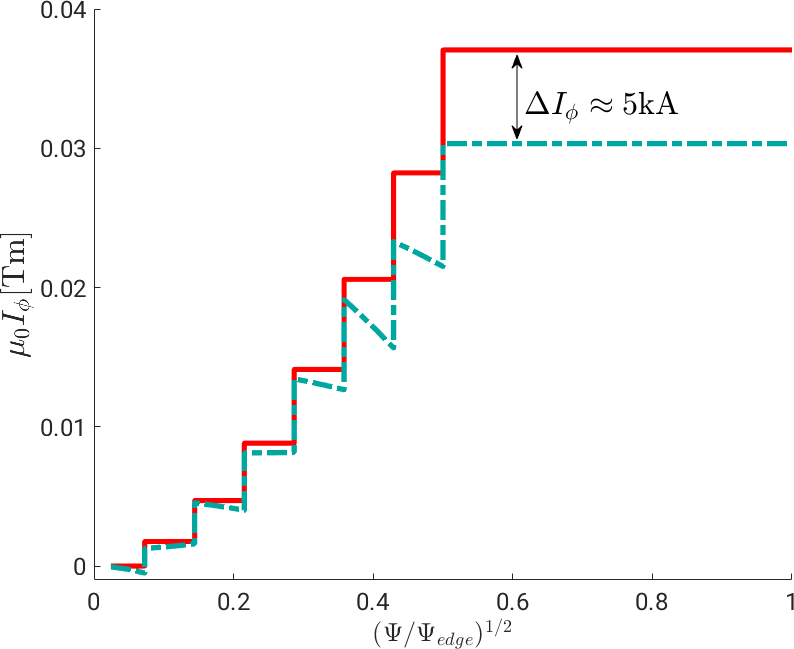}
    \caption{Total enclosed toroidal current as a function of the normalized square-root of the toroidal flux. In red: initial equilibrium and in blue: $\{I^v_{\phi,l}\}_{l=\{1,\ldots,7\}}$ optimized equilibrium. Other optimized equilibria have the same toroidal current profile as the initial equilibrium, and are thus not plotted. }
    \label{fig:Iprofile}
\end{figure}

The rotational transform profile is plotted on Figure \ref{fig:iota_profile} and the total enclosed toroidal current on Figure \ref{fig:Iprofile}. Again, the rotational transform is of the same order as for the un-optimized case. Regarding the optimized current profile, the total injected current is $\Delta I_\phi=-5.2$kA, less than $20\%$ of the initial total enclosed toroidal current.

\subsection{Convergence and computation time}

\begin{figure}
    \centering
    \includegraphics[width=.45\textwidth]{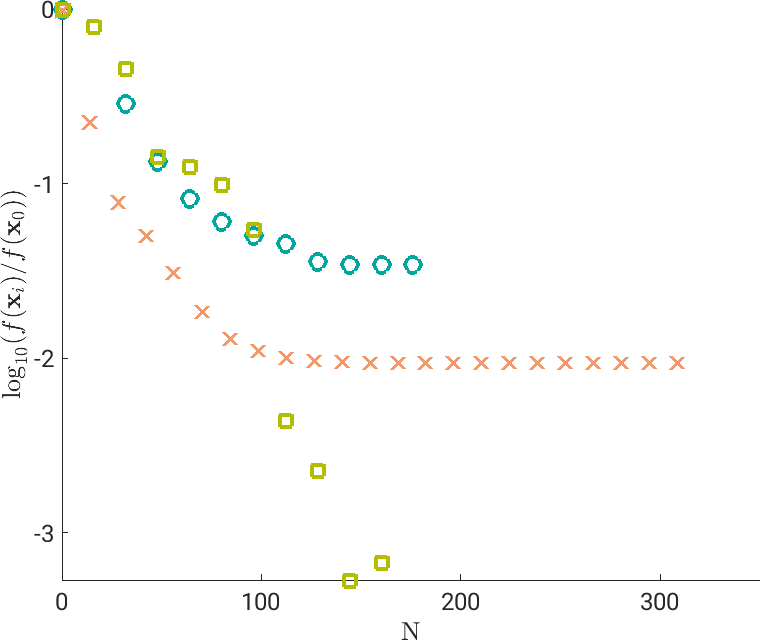}
    \caption{Logarithmic plot of the normalized objective function as a function of the number of SPEC calculations. Values evaluated while taking finite differences to evaluate the jacobian of the objective function are not shown.  Orange crosses: optimization of $\Gamma_w$. Blue circles: optimization of $\{I^v_{\phi,l}\}_{l=\{1,\ldots,7\}}$. Green squares: optimization of $\{V_{mn}\}$.}
    \label{fig:objective_function}
\end{figure}

The normalized value of the objective function as a function of the number of iterations is plotted for each optimization in Figure \ref{fig:objective_function}. We see that the optimization of $\{I^v_{\phi,l}\}_{l=\{1,\ldots,7\}}$ saturates at a larger value than the optimization of $\{V_{mn}\}$, despite optimizing the same objective function (the same residues were selected). 

The optimization was run in parallel on $2n+1$ cores of Intel Broadwell processors at 2.6GHz, where $n$ is the number of degrees of freedom of the optimization. Each core computed a different SPEC equilibrium when evaluating the finite difference estimate of the objective function gradient. The $\{V_{mn}\}$ optimization required $165$ equilibrium calculations, the $\{I^v_{\phi,l}\}_{l=\{1,\ldots,7\}}$ optimization  $187$ and the $\Gamma_w$ optimization $309$. The total CPU time for execution was respectively $\sim41$ days, $\sim10$ days and $\sim21$ days for a total wall-clock time of respectively $\sim 65$h, $\sim 17$h and $\sim 39$h. As expected, fixed-boundary optimizations were faster. 

Note that the presented optimizations did not take advantage of the full parallelization of SPEC --- only a single core computed each SPEC equilibrium, while SIMSOPT allows the user to use $M$ CPUs on $N$ SPEC instances, which would speedup the computation greatly. Nevertheless, our optimizations show that the total time required to perform a SPEC optimization is small enough to be considered in more advanced stellarator optimizations.

\section{Conclusion}

In this paper, we showed the first fixed- and free-boundary, multi-volume, finite $\beta$ SPEC equilibrium optimizations of a classical stellarator using the SIMSOPT framework. The objective function was constructed from the Greene's residues of selected rational surfaces. Different parameter spaces were considered: either the boundary of a perfectly conducting wall surrounding the plasma, or the enclosed toroidal current profile, or the vacuum field produced by the coils were optimized. 

In all three optimizations, it was possible to reduce the objective function significantly, which in turn translated to a reduction of the targeted magnetic island width. With the exception of the case of the coil optimization, all optimized states have a larger volume occupied by nested flux surfaces, which is beneficial for confinement. In the case of the coil optimization, additional rationals emerged and their related residues were not optimized, since they were not included in the objective function, but the islands present in the initial un-optimized equilibrium, were reduced in size.

Different measures of the magnetic field integrability are currently being considered to overcome the shortcomings of Greene's residues. Indeed, as observed in the coil optimization, Greene's residue is a local measure, which requires input from the user --- any rational surfaces emerging during the optimization remains undetected. Ideally, a global measure is thus required. The volume occupied by chaotic field lines, measured from the fractal dimension of their Poincare map \cite{Loizu2017} could be a possible solution. 

In the work by \citet{Landreman2021a}, it has been shown that SPEC could be coupled to VMEC in order to achieve an optimization in a vacuum, where both quasi-symmetry and nested flux surfaces could be obtained. The obvious next step is then to perform a combined SPEC-VMEC finite-$\beta$ optimization for good magnetic surfaces as well as other metrics, such as quasi-symmetry. This will require to deviate significantly from a classical stellarator geometry, as opposed to what has been presented in this paper. This will be the topic of future investigations.

\begin{acknowledgments}
The authors would like to acknowledge the support from the SIMSOPT development team, and thank Z. Qu, B. Medasani and C. Zhu for useful discussions. 
This work has been carried out within the framework of the EUROfusion Consortium and has received funding from the Euratom research and training programme 2014 - 2018 and 2019 - 2020 under grant agreement No 633053. The views and opinions expressed herein do not necessarily reflect those of the European Commission. This work was supported by a grant from the Simons Foundation
(560651, ML). BM and CZ are supported by the U.S. Department of Energy under Contract
No. DE-AC02-09CH11466 through the Princeton Plasma Physics Laboratory.
\end{acknowledgments}

\section*{Data Availability Statement}

The data that support the findings of this study are available from the corresponding author upon reasonable request.

\section*{References}

\bibliography{references.bib}

\end{document}